\documentclass[twocolumn,aps,superscriptaddress,showpacs]{revtex4}
\usepackage{amssymb}
\usepackage{amsmath}
\usepackage{graphicx}
\usepackage[normalem]{ulem}
\usepackage[dvips]{color}

\renewcommand\sout{\bgroup \color{red} \ULdepth=-.5ex \ULset}

\renewcommand\sout{\bgroup \color{red} \ULdepth=-.5ex \ULset}

\begin{document}

\title{Quark stars under strong magnetic fields}
\author{Peng-Cheng Chu}
\affiliation{Department of Physics and Astronomy and Shanghai Key Laboratory for Particle
Physics and Cosmology, Shanghai Jiao Tong University, Shanghai 200240, China}
\author{Lie-Wen Chen\footnote{%
Corresponding author (email: lwchen$@$sjtu.edu.cn)}}
\affiliation{Department of Physics and Astronomy and Shanghai Key Laboratory for Particle
Physics and Cosmology, Shanghai Jiao Tong University, Shanghai 200240, China}
\affiliation{Center of Theoretical Nuclear Physics, National Laboratory of Heavy Ion
Accelerator, Lanzhou 730000, China}
\author{Xin Wang}
\affiliation{Department of Physics and Astronomy and Shanghai Key Laboratory for Particle
Physics and Cosmology, Shanghai Jiao Tong University, Shanghai 200240, China}
\date{\today}

\begin{abstract}
Within the confined isospin- and density-dependent mass model, we study
the properties of strange quark matter (SQM) and quark stars (QSs) under strong magnetic
fields. The equation of state of SQM under a constant magnetic field is obtained
self-consistently and the pressure perpendicular to the magnetic field is shown
to be larger than that parallel to the magnetic field, implying that the properties
of magnetized QSs generally depend on both the strength and the
orientation of the magnetic fields distributed inside the stars. Using a density-dependent
magnetic field profile which is introduced to mimic the magnetic field strength
distribution in a star, we study the properties of static spherical QSs by
assuming two extreme cases for the magnetic field orientation in the stars, i.e.,
the radial orientation in which the local magnetic fields are along the radial
direction and the transverse orientation in which the local magnetic fields are
randomly oriented but perpendicular to the radial direction. Our results indicate
that including the magnetic fields with radial (transverse) orientation can significantly
decrease (increase) the maximum mass of QSs, demonstrating the importance
of the magnetic field orientation inside the magnetized compact stars.
\end{abstract}

\maketitle

\section{Introduction}

The compact stars provide a unique astrophysical testing ground to explore
the nature of matter under extreme conditions~\cite{Gle00,Web99}. Neutron
stars (NSs) are a class of densest compact stars in the universe. In the
interior of NSs, the baryon number density can reach or even be larger than
about $6$ times normal nuclear matter density and thus hyperons, meson
condensations and even quark matter may be present there. The study of NSs
has provided us important information about the equation of state (EOS) of
neutron-rich nuclear matter~\citep{Lat04,Ste05}. Theoretically, NSs may be
converted to (strange) quark stars (QSs), which is made purely of deconfined
$u$, $d$, and $s$ quark matter (with some leptons due to charge neutrality and $\beta $-equilibrium),
i.e., strange quark matter (SQM)~\cite{Bom04,Sta07,Her11}.
The possible existence of QSs is one of the most intriguing aspects of
modern astrophysics and has important implications for the strong
interaction matter at high baryon densities, especially the
properties of SQM that essentially determine the structure of QSs.
In terrestrial laboratories, the properties of SQM can be explored by heavy
ion collisions, e.g., the beam-energy scan program at RHIC as well as the
experiments planned in the Facility for Antiproton and Ion Research (FAIR)
at GSI and the Nuclotron-based Ion Collider Facility (NICA) at JINR, which
aim to give a detailed picture of the QCD phase structure, especially to
locate the so-called QCD critical point~\cite{Ste98}. These studies on SQM have
become nowadays one of the fundamental issues in nuclear physics,
astrophysics and cosmology.

Theoretically, according to the Bodmer-Witten-Terazawa
hypothesis~\cite{Bodmer71,Terazawa79,Witten84}, SQM might be the true ground
state of QCD matter (i.e., the strong interaction matter) and is absolutely
stable. The properties of SQM in QSs cannot be calculated directly by either the
{\it ab initio} Lattice QCD or the perturbative QCD (pQCD) because of the
difficulty in treating the finite baryon chemical potential or the low energy scale,
and thus a number of phenomenological models have been proposed to explore
the properties of SQM, such as MIT bag model~\cite{Cho74,Far84,Alc86,Alf05,Web05},
the Nambu-Jona-Lasinio (NJL) model~\citep{Rehberg96,Han01,Rus04,Men06}, the pQCD
approach~\cite{Fre77,Fra01,Fra05,Kur10}, the Dyson-Schwinger
approach~\cite{Rob94,Zon05,Qin11}, the confined-density-dependent-mass
(CDDM) model~\cite{Fow81,Cha89,Cha91,Ben95,Pen99,Peng00,Peng08,Li11},
and the quasi-particle model~\citep{Sch97,Sch98}. Within an isospin-extended
version of the CDDM model, i.e., the confined isospin- and density-dependent mass
(CIDDM) model~\citep{Chu2014} in which the quark confinement is modeled by the density- and
isospin-dependent quark masses, it has been shown recently that QSs provide an excellent
astrophysical laboratory to explore the properties of SQM, especially the
quark matter symmetry energy.

In the work of Ref.~\citep{Chu2014}, it has been assumed that the magnetic field
effects can be neglected in QSs. An important aspect of the compact star physics 
is that compact stars could be endowed with strong magnetic fields. Large magnetic field
strength of $B\sim10^{14}$ G has been estimated at the surface of compact
stars~\citep{Wol64,Mihara90,Chan92}. The magnetic field strength may reach as
large as $B\sim10^{18}$ G in the core of compact stars~\cite{Lai91}. In the work by
Ferrer~{\it et al.}~\citep{Ferrer10}, the estimated magnetic field strength in the
core of the self-bound QSs can even reach about $10^{20}$ G. In such tremendous magnetic
fields, the spatial rotational ($\mathcal{O}(3)$) symmetry will break and one must
consider the pressure anisotropy of the system~\cite{Ferrer10,Isa11,Isa12,Isayev13}.
Furthermore, in order to describe the spatial distribution of the magnetic field
strength in compact stars, people usually introduce a density-dependent magnetic
field profile~\citep{Ban97,Ban98}. Therefore, it is
interesting and important to study the effects of the spatial distribution of the
magnetic field strength and orientation on the properties of compact stars. These
studies are critical for accurately determining the properties (e.g., EOS) of
dense matter by comparing the model results with the astrophysical observations of
compact stars. As a matter of fact, it is still controversial about if the inclusion
of the magnetic fields can enhance or reduce the maximum mass of the compact
stars~\citep{Ban97,Ban98,Menezes09,Ryu10,Ryu12,Bro00,Car01,Pau11,Dex14,Cas14,Hou14}.
The main motivation of the present work is to explore the properties of SQM and QSs
under strong magnetic fields. We demonstrate that both the strength distribution and
the orientation of the magnetic fields inside the QSs are important for understanding 
the properties of the QSs, and depending on the magnetic field orientation, the 
maximum QS mass can be either enhanced or reduced.

The paper is organized as follows. We describe in Sec.~\ref{model} the
theoretical models and methods used in the present paper, and then present
the results and discussions in Sec.~\ref{result}. Finally, a conclusion
is given in Sec.~\ref{summary}.

\section{Models and methods}
\label{model}

\subsection{The confined isospin- and density-dependent mass model}

The confined isospin- and density-dependent mass (i.e., the CIDDM) model~\cite{Chu2014}
is an extended version of the CDDM model~\cite{Fow81,Cha89,Cha91,Ben95,Pen99,Peng00,Peng08,Li11}
for quark matter by introducing the isospin dependence of the quark equivalent
mass. In the CIDDM model, the quark confinement is modeled by the density- and
isospin-dependent quark masses. Particularly, the equivalent quark mass in
isospin asymmetric quark matter with baryon number density $n_B$ and isospin
asymmetry $\delta $ is expressed as
\begin{eqnarray}
m_q &=& m_{q_0} + m_{I} + m_{iso} \notag \\
&=& m_{q_0} + \frac{D}{{n_B}^z} - \tau_q \delta {D_I}n_B^{\alpha}e^{-\beta n_B},
\label{mqiso}
\end{eqnarray}
where $m_{q0}$ is the quark current mass, $m_I = \frac{D}{{n_B}^z} $ reflects
the flavor-independent quark interactions in quark matter, and
$m_{iso} = - \tau_q \delta {D_I}n_B^{\alpha}e^{-\beta n_B}$ represents the
isospin dependent quark interactions in quark matter. For $m_I = \frac{D}{{n_B}^z} $,
the constant $z$ is the quark mass scaling parameter and the constant $D$ is a
parameter determined by stability arguments of SQM. For
$m_{iso} = - \tau_q \delta {D_I}n_B^{\alpha}e^{-\beta n_B}$, the constants $D_I$,
$\alpha $ and $\beta $ are parameters determining the isospin dependence of quark
effective interactions in quark matter, $\tau_q $ is the isospin quantum number
for quarks and we set $\tau_q = 1$ for $q=u$ ($u$ quarks), $\tau_q = -1$ for $q=d$
($d$ quarks), and $\tau_q = 0$ for $q=s$ ($s$ quarks). As usual~\cite{DiT06,Pag10,DiT10,Sha12},
the isospin asymmetry is defined as
\begin{equation}
\delta = 3\frac{n_d-n_u}{n_d+n_u},
\label{delta}
\end{equation}
which equals to $-n_3/n_B$ with the isospin density $n_3 = n_u-n_d$ and
$n_B = (n_u+n_d)/3$ for two-flavor $u$-$d$ quark matter. Especially, one has
$\delta = 1$ ($-1$) for quark matter converted by pure neutron (proton) matter
according to the nucleon constituent quark structure, consistent with the
conventional definition for nuclear matter, i.e., $\frac{n_n -n_p}{n_n +n_p}=-n_3/n_B$.

From Eq. (\ref{mqiso}), one can see that the quark confinement condition
$\lim_{n_B\to0}m_q=\infty$ can be guaranteed if $z > 0$ and $\alpha \ge 0$.
Furthermore, if $\beta >0$, one then has $\lim_{n_B \to \infty} m_{iso} = 0$
and thus the asymptotic freedom $\lim_{n_B\to \infty} m_q=m_{q0}$ is satisfied.
For two-flavor $u$-$d$ quark matter, the chiral symmetry is restored at high
density due to $\lim_{n_B\to\infty}m_q = 0$ if the current masses of $u$ and $d$
quarks are neglected. In addition, the equivalent quark mass in Eq.~(\ref{mqiso})
also satisfies the exchange symmetry between $u$ and $d$ quarks which is required
by isospin symmetry of the strong interaction. Therefore, the phenomenological
parametrization form of the isospin dependent equivalent quark mass in
Eq.~(\ref{mqiso}) is quite general and respects the basic features of QCD.
For more details about the CIDDM model, the reader is referred to Ref.~\cite{Chu2014}.

As demonstrated in Ref.~\cite{Chu2014}, including the isospin dependent quark effective
interactions $m_{iso} $ in the equivalent quark mass can significantly influence
the quark matter symmetry energy as well as the properties of SQM and QSs. In
particular, the most recently discovered large mass pulsar PSR J0348+0432 with
a mass of $2.01\pm0.04M_{\odot}$ can be described as a QS within the CIDDM model
if the two-flavor $u$-$d$ quark matter symmetry energy is large enough and the
value of the quark mass scaling parameter $z $ is selected appropriately. For
instance, the parameter set DI-85 ($z=1.8$), for which we have
$D_I = 85$ MeV$\cdot$fm$^{3\alpha}$, $\alpha=0.7$, $\beta=0.1$ fm$^3$, $D=22.922$
MeV$\cdot$fm$^{-3z}$, $z=1.8$, $m_{u0}=m_{d0}=5.5$ MeV and $m_{s0}=80$ MeV,
can predict a QS with a mass of $2.01M_{\odot }$, corresponding to the
measured center value for the pulsar PSR J0348+0432~\cite{Ant13}. The corresponding
radius of the predicted QS (with a mass of $2.01M_{\odot }$) is $9.98$ km,
the central baryon number density is $1.25$ fm$^{-3}$, and the surface
(zero-pressure point) baryon number density is $0.465$ fm$^{-3}$. In addition,
the strength of the two-flavor $u$-$d$ quark matter symmetry energy from DI-85 ($z=1.8$)
is about two times larger than that of the free quark gas or that predicted by the
conventional NJL model. In this work, we study the properties of SQM and QS's
under strong magnetic fields in the CIDDM model with DI-85 ($z=1.8$).

\subsection{SQM under a constant magnetic field}

The energy spectrum of a fermion (e.g., quarks and leptons) with
electric charge $q_i$ in an external constant magnetic field with strength $B$ can
be expressed as~\cite{Landau1965}
\begin{align}
E_{p,i}&=\sqrt{p_z^2+2\nu |q_i| B+m_i^2},&
\end{align}
where $p_z$ is the momentum in the $z$ direction (here the magnetic field
is assumed to be along the $z$ axis), $m_i$ is mass, and
$\nu=n+\frac{1}{2}-\frac{q_i}{|q_i|}\frac{s}{2}$ represents the Landau
levels with $n=0,1,2,3,...$ being the principal quantum number and
$s=\pm1$ denoting the spin (``$+1$'' for spin-up and ``$-1$'' for spin-down).
Similarly to the works in Refs.~\cite{Ban97,Ban98,Rabhi09,Wen12,Wen13,Isayev13},
we do not consider here the contributions from the anomalous magnetic moments
since they are not well understood for quarks in the deconfined condition 
and are insignificant for leptons~\cite{Duncan00}.

In this work, we study the thermodynamic properties of magnetized SQM at
zero temperature within the CIDDM model. The thermodynamic potential
of SQM under a constant magnetic field $B$ can then be expressed as
\begin{align}
\Omega=\sum\limits_{i=u,d,s,l}\Omega_i,
\end{align}
with
\begin{align}
\Omega_i=&-
\sum\limits_{\nu=0}^{\nu_{max}^i}\frac{g_i(|{q_i}|B)}{4\pi^2}\alpha_{\nu} \notag\int_{-\infty}^{\infty}\text{d}p_z [\mu_i^*-E_{p,i}]&\notag\\
=&-
\sum\limits_{\nu=0}^{\nu_{max}^i}\frac{g_i(|{q_i}|B)}{2\pi^2}\alpha_{\nu}\bigg\{\frac{1}{2}\mu_i^*
\sqrt{{\mu_i^*}^2-s_i(\nu,B)^2}&\notag\\ &-\frac{s_i(\nu,B)^2}{2}\ln\bigg[\frac{\mu_i^*+\sqrt{{\mu_i^*}^2-s_i(\nu,B)^2}}{s_i(\nu,B)}\bigg]\bigg\}.&
\end{align}
In the above expressions, $\Omega_i$ represents each particle contribution to the thermodynamic
potential, $i$ in the sum is for all flavors of quarks and leptons, and
$\alpha_{\nu}=2-\delta_{\nu,0}$. The degeneracy factor $g_i=3$ for quarks and
$g_i=1$ for leptons, and the Fermi energy for quarks and leptons is
\begin{eqnarray}
\mu_i^*=\sqrt{{k_{F,\nu}^i}^2+s_i(\nu,B)^2},
\end{eqnarray}
with $k_{F,\nu}^i$ being the Fermi momentum and $s_i(\nu,B)=\sqrt{m_i^2+2\nu|q_i|B}$.
The highest Landau level is defined as
\begin{eqnarray}
 \nu_{max}^i\equiv \text{int} \bigg[\frac{{\mu_i^*}^2-m_i^2}{2 |{q_i}|B}\bigg],
\end{eqnarray}
where int[$\cdot \cdot \cdot$] is the integer function.
The total energy density of the system is thus obtained as
\begin{align}
 \mathcal{E}_{tot} =&\Omega +\sum\limits_{i=u,d,s,l}{\mu_i^*}n_i\notag&\\
 = &\sum\limits_{i=u,d,s,l}\sum\limits_{\nu=0}^{\nu_{max}^i}\frac{g_i(|{q_i}|B) }{4\pi^2}\alpha_{\nu} \notag\int_{-\infty}^{\infty}\text{d}p_z E_{p,i}+\frac{B^2}{2}& \notag\\= & \sum\limits_{i=u,d,s,l} \sum\limits_{\nu=0}^{\nu_{max}^i}\frac{g_i(|{q_i}|B) }{4\pi^2}\alpha_{\nu}\bigg\{\mu_i^* \sqrt{{\mu^*_i} ^2-s_i(\nu,B)^2} &  \notag\\
 &+s_i(\nu,B)^2\ln{\bigg [\frac{\mu_i^*+\sqrt{{\mu^*_i}^2-s_i(\nu,B)^2}}{s_i(\nu,B)}\bigg ]}\bigg\}+\frac{B^2}{2},&
\end{align}
where the term $B^2/2$ comes from the magnetic field contribution and $n_i$ is the
number density of quarks and leptons given by
\begin{eqnarray}
n_i=\frac{g_i|q_i|B}{2\pi^2}\sum\limits_{\nu=0}^{\nu_{max}^i}(2-\delta_{\nu},0)k_{F,\nu}^i.
\end{eqnarray}

We note that for the cases we consider in this work, muons will not appear in SQM
due to the small chemical potential of electrons. The electric charge neutrality
condition of SQM can thus be written as
\begin{eqnarray}
&&\frac{2}{3}n_u=\frac{1}{3}n_d+\frac{1}{3}n_s+n_e.
\end{eqnarray}
For SQM, we assume it is neutrino free and the $\beta$-equilibrium condition
in SQM can then be expressed as
\begin{eqnarray}
&&\mu_u+\mu_e=\mu_d=\mu_s,
\end{eqnarray}
where $\mu_i$ ($i=u$, $d$, $s$ and $e^-$) represents the particle chemical potential.
For quarks, the chemical potential in SQM can be obtained as
\begin{eqnarray}
\mu_i =\frac{d\mathcal{E}_{tot}}{d n_i}=\mu_i^*+\sum\limits_{j=u,d,s}\frac{\partial{\Omega_j}}{\partial m_j}\frac{\partial m_j}{\partial n_i}.
\label{mu}
\end{eqnarray}
One can see from Eq.~(\ref{mu}) that, owing to the density dependence
of the equivalent quark mass, there are additional terms in the chemical
potential compared to the case of free Fermi gas. Therefore, the chemical
potential of $u$, $d$ and $s$ quarks in SQM can be obtained, respectively, as
\begin{align}
\mu_u&=\mu_u^*+D_I n_B^\alpha e^{-\beta n_B}\bigg [\frac{\partial \Omega_u}{\partial m_u}-\frac{\partial\Omega_d}{\partial m_d}\bigg ]\frac{6n_d}{(n_u+n_d)^2}+\mu_{den},&\\
\mu_d&=\mu_d^*+D_I n_B^\alpha e^{-\beta n_B}\bigg [\frac{\partial \Omega_d}{\partial m_d}-\frac{\partial\Omega_u}{\partial m_u}\bigg ]\frac{6n_u}{(n_u+n_d)^2}+\mu_{den},&\\
\mu_s&=\mu_s^* + \mu_{den},&
\end{align}
with
\begin{eqnarray}
\frac{\partial \Omega_f}{\partial m_f}&=&\frac{3}{2\pi^2}\sum\limits_{\nu=0}^{\nu_{max}^f}\alpha_\nu(|{q_f|}Bm_f)\times \notag\\
&&\ln\bigg [\frac{k^{f}_{F,\nu}+\sqrt{{k^{f}_{F,\nu}}^2+2\nu|q_f|B+m_f^2}}{\sqrt{2\nu|{q_f}|B+m_f^2}}\bigg ],
\end{eqnarray}
and
\begin{eqnarray}
\mu_{den}&=&\frac{1}{3}\sum\limits_{j=u,d,s}\frac{3}{2\pi^2}\sum\limits_{\nu=0}^{\nu_{max}^j}\alpha_\nu(|{q_j}|B)m_j\times \notag\\
&&\ln\bigg [\frac{k^{j}_{F,\nu}+\sqrt{{k^{j}_{F,\nu}}^2+2\nu|{q_j}|B+m_j^2}}{\sqrt{2\nu|{q_j}|B+m_j^2}}\bigg ]\times \notag\\
&&\bigg\{-\frac{zD}{n_B^{(1+z)}}-\tau_jD_I\delta[\alpha n_B^{\alpha-1}-\beta n_B^\alpha]e^{-\beta n_B}\bigg\}.
\end{eqnarray}
The chemical potential of leptons can be written as
\begin{eqnarray}
{\mu_l}=\sqrt{{k_{F,\nu}^l}^2+s_l(\nu,B)^2}.
\end{eqnarray}

For SQM under a constant magnetic field, the $\mathcal{O}(3) $ rotational symmetry
is broken and the pressure of the system becomes anisotropic, i.e., it is split into the
longitudinal pressure $P_\parallel$ which is parallel to the magnetic field and
the transverse pressure $P_\perp$ which is perpendicular to the magnetic field.
The expressions of $P_\parallel$ and $P_\perp$ for a magnetized fermion system can
be written as~\cite{Ferrer10}
\begin{eqnarray}
P_\parallel&=&\sum\limits_i \mu_i n_i - \mathcal{E}_{tot}, \label{Plong} \\
P_\perp&=&\sum\limits_i \mu_i n_i - \mathcal{E}_{tot}+B^2-M B, \label{Pperp}
\end{eqnarray}
where $M$ is the system magnetization. It is interesting to see that the
longitudinal pressure $P_\parallel$ satisfies the Hugenholtz-Van Hove (HVH)
theorem~\cite{HVH58} while the transverse pressure $P_\perp$ has extra contributions
from the magnetic field. This feature will lead to the fact that the
zero-pressure point density coincides with the density at the minimum of
the energy per baryon for $P_\parallel$ but not for $P_\perp$, as will be
shown later.

In the CIDDM model, the longitudinal and transverse pressures of the system
under a constant magnetic field can thus be expressed, respectively, as
\begin{eqnarray}
P_\parallel&=&-\sum\limits_{i=u,d,s,l} \Omega_i +\sum\limits_{i,j=u,d,s}\frac{\partial \Omega_j}{\partial m_j} \frac{\partial m_j}{\partial n_i}n_i-\frac{B^2}{2}, \label{Plong2}\\
P_\perp&=&-\sum\limits_{i=u,d,s,l} \Omega_i +\sum\limits_{i,j=u,d,s}\frac{\partial \Omega_j}{\partial m_j} \frac{\partial m_j}{\partial n_i}n_i+\frac{B^2}{2}-M B, \notag\\ \label{Pperp2}
\end{eqnarray}
where the system magnetization $M$ is given by
\begin{align}
M=-\partial{\Omega}/\partial{B}=\sum\limits_{i=u,d,s,l} M_i,
\end{align}
with
\begin{align}
M_i=-\frac{g_i|{q_i}|}{2\pi^2}\sum\limits_{\nu=0}^{\nu_{max}^i}(2-\delta_{\nu 0})\int_0^{k_{F,\nu}^i}\bigg[\frac{\nu|{q_i}|B}{\epsilon_\nu^i}+\epsilon_\nu^i-\mu_i^*\bigg]dk_z,
\end{align}
and $\epsilon_\nu^i=\sqrt{k_z^2+m_i^2+2\nu|{q_i}|B} $. It should be emphasized that
the longitudinal and transverse pressures in Eqs.~(\ref{Plong2}) and (\ref{Pperp2})
include the contributions from the magnetic fields. In particular, one can see
that the magnetic energy density term $B^2/2$ contributes oppositely to the longitudinal
and transverse pressures under a constant magnetic field, and it decreases the
former while increases the latter, which will lead to a tremendous difference
between the longitudinal and transverse pressure when the magnetic field is very strong.

\subsection{Density-dependent magnetic fields in quark stars}

It is generally believed that the magnetic field strength in
the core of compact stars should be much larger than that at
the surface, and a density-dependent magnetic field profile is
usually introduced to describe this behavior for the spatial
distribution of the magnetic field strength in the compact
stars~\cite{Ban97}. In the present work, we use the following
popular parametrization for the density-dependent magnetic field
profile in QSs~\citep{Ban97,Ban98,Menezes09,Ryu10,Ryu12}
\begin{eqnarray}
B=B_{surf}+B_0[1-\exp{(-\beta_0(n_B/n_0)^\gamma)}], \label{BrhoEq}
\end{eqnarray}
where $B_{surf}$ is the magnetic field strength at the surface of
compact stars and its value is fixed at $B_{surf} = 10^{15}$G in this work,
$n_0 = 0.16$ fm$^{-3}$ is the normal nuclear matter density, $B_0$
is a parameter with dimension of $B$, $\beta_0 $ and $\gamma $ are
two dimensionless parameters that control how exactly the magnetic field
strength decays from the center to the surface.

Besides the spatial distribution of the magnetic field strength
in the compact stars, the orientation of the magnetic fields
is also expected to be important for the structure of the compact
stars since the pressure (including the contribution from the magnetic
fields) may become significantly anisotropic under strong magnetic
fields. Consequently, the gravitational field in magnetized stars is
no longer spherically symmetric due to the pressure anisotropy and the well-known
Tolman-Oppenheimer-Volkoff (TOV) equations~\citep{Oppenheimer39}
generally cannot be applied to calculate the structure of the
magnetized compact stars since they are only valid for spherically
symmetric compact stars.

Since there is no empirical information on the magnetic field orientation
inside the compact stars, in the present work, we assume two extremely
special cases for the orientation of the magnetic fields inside the compact
stars: one is that the local magnetic fields are along the radial direction
(denoted as ``radial orientation''), and the other is that the local magnetic
fields are perpendicular to the radial direction but randomly oriented in
the plane perpendicular to the radial direction (denoted as ``transverse
orientation''). It should be mentioned that the magnetic fields pass
through the centra of the spherical compact stars for the radial orientation.
In these two extreme cases, the pressure distribution inside the static
compact stars can be considered to be spherically symmetric and thus the
gravitational field as well as the static compact stars are spherically
symmetric too. For these two extreme cases for the orientation of the
magnetic fields, one thus can calculate the structure of the static
magnetized compact stars by solving the following TOV equations:
\begin{align}
\frac{dM(r)}{dr}=&4\pi r^2 \epsilon(r), \\
\frac{dp(r)}{dr}=&-\frac{G\epsilon(r)M(r)}{r^2}\bigg[1+\frac{p(r)}{\epsilon(r)}\bigg]\bigg[1+\frac{4\pi p(r)r^3}{M(r)}\bigg]\times \notag \\
&\bigg[1-\frac{2GM(r)}{r}\bigg]^{-1},
\end{align}
where $M(r)$ is the total mass inside the sphere of radius $r$, $\epsilon(r)$
is the corresponding energy density (including the magnetic field contribution),
$p(r)$ is the corresponding (radial) pressure (including the magnetic field
contribution), and $G$ is Newton's gravitational constant.

We would like to point out that the radial and transverse orientations have
been assumed to be inside the compact stars and the magnetic field orientation
outside the compact stars or close the surface of the compact stars may change
and become nonspherically symmetric. On the other hand, it should be mentioned that around
the surface of compact stars, the magnetic field strength relatively is quite
weak ($\sim10^{15}$ G) and the pressure is very small, and thus the magnetic
fields around the surface is not important for the structure of QSs. For more
general cases of magnetic field orientations and/or distributions in the magnetized
compact stars where the spherical symmetry is broken, Einstein field equations
should be solved self-consistently to calculate the structure of the compact
stars, and this is beyond the scope of the present work and yet to be
constructed.

\section{Results and discussions}
\label{result}

\subsection{EOS of SQM under a constant magnetic field}

\begin{figure}[tbp]
\includegraphics[scale=0.85]{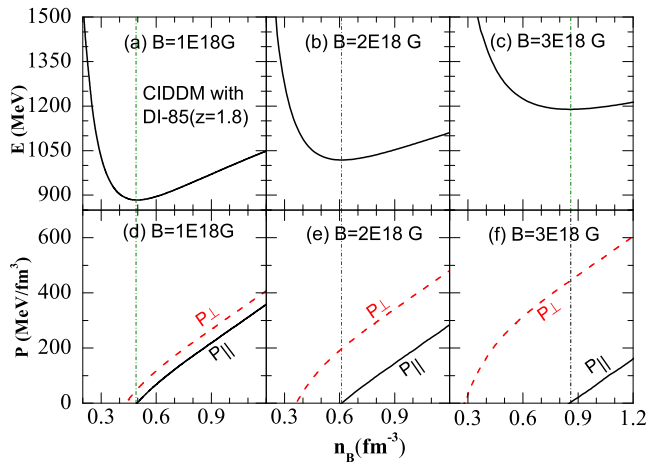}
\caption{(Color online) Energy per baryon and the corresponding longitudinal
and transverse pressures as functions of the baryon density for SQM under constant
magnetic fields with strengthes of $B = 1\times10^{18}$ G, $2\times10^{18}$ G and $3\times10^{18}$ G
within the CIDDM model with DI-85 (z=1.8).}
\label{EOSB}
\end{figure}
We first present the results for the properties of SQM under a constant magnetic
field. Using the CIDDM model with DI-85 (z=1.8), we show in Fig.~\ref{EOSB} the
energy per baryon and the corresponding longitudinal and transverse pressures as
functions of the baryon density for SQM under constant magnetic fields with three
strengthes of $B = 1\times10^{18}$ G, $2\times10^{18}$ G and $3\times10^{18}$ G.
One can see that for all the three values of the magnetic field strength $B$, the
density at the minimum of energy per baryon is exactly equal to the zero point
density of the longitudinal pressure $P_\parallel$, which is consistent with HVH
theorem (as shown in Eq.~(\ref{Plong})) and the thermodynamic self-consistency as
in the case without magnetic fields~\cite{Chu2014}. Furthermore, it is seen that
the density at the minimum of energy per baryon increases with the magnetic field
strength, i.e., it varies from $0.49$ fm$^{-3}$ to $0.61$ fm$^{-3}$ and then to
$0.86$ fm$^{-3}$ when $B$ changes from $1\times10^{18}$ G to $2\times10^{18}$ G
and then to $3\times10^{18}$ G.

In addition, one can also see from Fig.~\ref{EOSB} that at a fixed density, the
transverse pressure $P_\perp$ increases while the longitudinal pressure $P_\parallel$
decreases with the increment of the magnetic field strength $B$, leading to a
clear splitting between $P_\perp$ and $P_\parallel$ under the constant magnetic
fields. This pressure splitting (i.e., anisotropy) rapidly increases with $B$. The pressure anisotropy under
strong magnetic fields is due to the additional terms of $B^2$ and magnetization of
the system appeared in $P_\perp$ as shown in Eqs.~(\ref{Plong}) and (\ref{Pperp}).

\begin{figure}[tbp]
\includegraphics[scale=1.2]{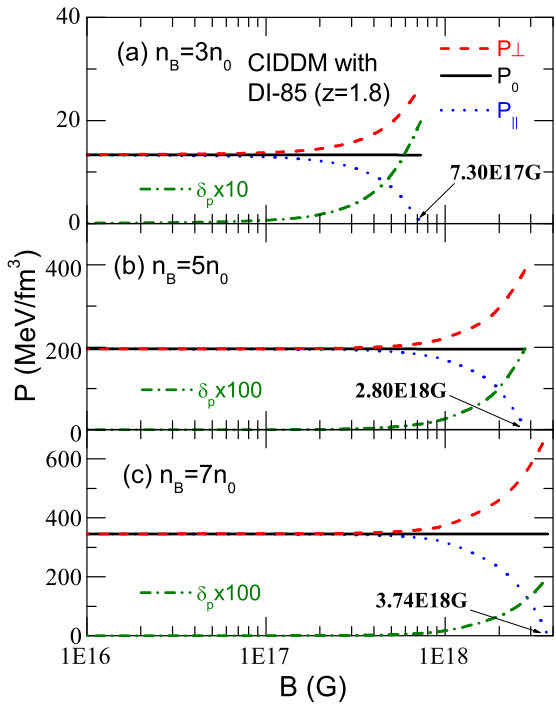}
\caption{(Color online) Transverse and longitudinal pressures together with the
pressure splitting factor $\delta_p $ as functions of the magnetic field strength
$B$ for SQM at baryon number densities of $n_{B} = 3n_0$, $5n_0$, and
$7n_0$ within the CIDDM model with DI-85 (z=1.8). The corresponding
pressures at $B=0$ ($P_0$) are also included for comparison.}
\label{PTLConstB}
\end{figure}

In order to quantitatively describe the pressure anisotropy under
strong magnetic fields, one can define a normalized pressure splitting
factor as
\begin{eqnarray}
\delta_p=\frac{P_{\perp}-P_{||}}{(P_{\perp}+P_{||})/2}.
\end{eqnarray}
From this definition, one has $\delta_p =0$ if there is no splitting
between $P_\perp$ and $P_\parallel$, and $\delta_p =2$ for the extremely
anisotropic case with $P_\parallel = 0$. Shown in Fig.~\ref{PTLConstB}
are the transverse and longitudinal pressures together with the pressure
splitting factor $\delta_p $ as functions of the magnetic field strength
$B$ for SQM at three baryon number densities of $n_{B} = 3n_0$,
$5n_0$, and $7n_0$ within the CIDDM model with DI-85 (z=1.8). For
comparison, the corresponding pressures at $B=0$ are also included
Fig.~\ref{PTLConstB}. We would like to point out that the central baryon
density in QSs is roughly around $7n_0$, and $n_{B} = 3n_0$
and $5n_0$ are two typical values of baryon density in QSs.

One can see from Fig.~\ref{PTLConstB} that, when the magnetic field strength
is larger than a certain value of $B_m$ below which the magnetic field effects
on the pressure are essentially negligible (with $\delta_p \le 5\%$), the
transverse pressure $P_\perp$ increases rapidly while the longitudinal pressure
$P_\parallel$ decreases rapidly with increment of $B$, leading to a rapid enhancement
of $\delta_p$. When the magnetic field strength $B$ further increases and reaches
a critical value of $B_c$, the $P_\parallel$ becomes to zero (and thus $\delta_p = 2$).
When the magnetic field strength is larger than $B_c$, the $P_\parallel$ 
becomes negative and thus the system becomes unstable. Therefore, $B_c$ is the
largest magnetic field strength that a stable SQM in QSs can have. In
addition, it is seen from Fig.~\ref{PTLConstB} that the values of $B_m$ and
$B_c$ depend on the baryon density, and particularly we have $B_m \approx 1.5\times10^{17}$ G
and $B_c \approx 7.30\times10^{17}$ G for $n_{B} = 3n_0$, $B_m \approx 4.5\times10^{17}$ G and
$B_c \approx 2.80\times10^{18}$ G for $n_{B} = 5n_0$, and $B_m \approx 6.0\times10^{17}$ G and
$B_c \approx 3.74\times10^{18}$ G for $n_{B} = 7n_0$. For a magnetized compact star, it
is thus important to ensure $B<B_c$ for all matter inside the compact stars.

\subsection{Quark stars under density dependent magnetic fields}

\begin{figure}[tbp]
\includegraphics[scale=0.5]{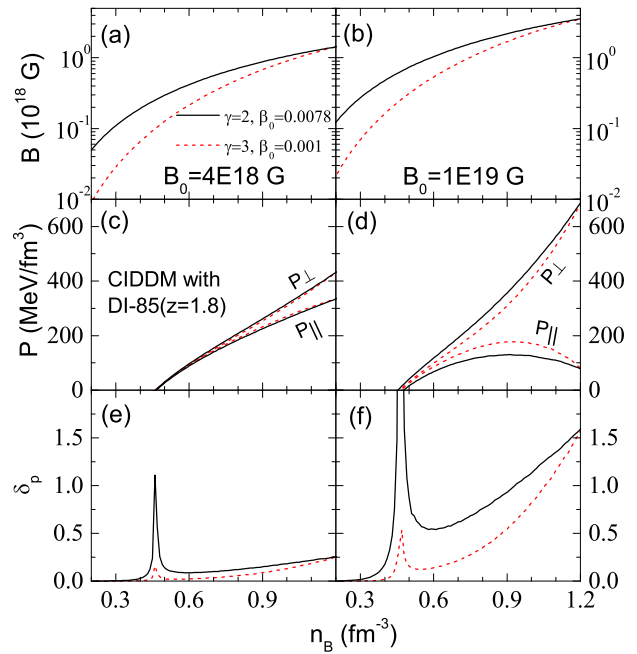}
\caption{(Color online) Baryon density dependence of the magnetic field
strength $B$, longitudinal and transverse pressures as well as the pressure
splitting factor $\delta_p$ for SQM in QSs using the slow B-profile (solid lines)
and fast B-profile (dashed lines) within the CIDDM model with DI-85 (z=1.8).
$B_0 = 4\times10^{18}$ G (left panels) and $1\times10^{19}$ G (right panels) are
considered.}
\label{Brho}
\end{figure}

As mentioned earlier, the magnetic field strength is generally believed
to be varied inside the magnetized compact stars, and a density-dependent
magnetic field profile of Eq.~(\ref{BrhoEq}) is usually introduced to mimic
the magnetic field strength distribution inside the stars. Due to our poor
knowledge on the magnetic field strength distribution inside the stars,
we consider in this work two sets of values for the dimensionless
parameters $\beta_0 $ and $\gamma $, i.e., the fast-varied magnetic
field profile with $\gamma =3$ and $\beta_0 = 0.001$ (denoted as
``fast B-profile'') and the slow-varied magnetic field profile
with $\gamma =2$ and $\beta_0 = 0.0078$ (denoted as ``slow B-profile'').
Using these two different magnetic field profiles, we show in Fig.~\ref{Brho}
the density dependence of the magnetic field strength $B$, the longitudinal
and transverse pressures as well as the corresponding pressure splitting
factor $\delta_p$ for SQM using $B_0 = 4\times10^{18}$ G and $B_0 = 1\times10^{19}$ G
within the CIDDM model with DI-85 (z=1.8).

From Fig.~\ref{Brho}, one can see that the fast B-profile gives a stronger
density dependence of magnetic field strength, i.e., a faster decay for the
magnetic field strength from high densities (e.g., the core of compact stars)
to low densities (e.g., the surface of compact stars), than the slow
B-profile as expected. For the smaller value of $B_0$ ($B_0 = 4\times10^{18}$ G),
one can see that $P_\perp$ is larger than $P_\parallel$ at higher densities and
then they approach to zero almost at the same density of about $0.46$ fm$^{-3}$.
For this smaller value of $B_0$, the pressure splitting between $P_\perp$ and
$P_\parallel$ is not so big and we have $\delta_p = 0.25$ (i.e.,
$P_\parallel / P_\perp =0.78$) at $1.2$ fm$^{-3}$. The peak of $\delta_p$
around $0.46$ fm$^{-3}$ is due to the vanishing of $P_\parallel$ there,
which corresponds to the surface of QSs. In addition, it is seen that the
difference between the EOSs of SQM with the fast B-profile and the slow
B-profile is small for $B_0 = 4\times10^{18}$ G. On the other hand, for
$B_0 = 1\times10^{19}$ G, one can see that $P_\perp$ is significantly larger
than $P_\parallel$ at higher densities, and while $P_\perp$ always increases
with $n_B$, $P_\parallel$ decreases with $n_B$ when $n_B \ge 0.9$ fm$^{-3}$,
leading to a very big pressure splitting between $P_\perp$ and $P_\parallel$ at
higher densities, i.e., $\delta_p = 1.6$ (corresponding to
$P_\parallel / P_\perp =1/9$) at $1.2$ fm$^{-3}$. Therefore, the pressure could
be strongly anisotropic in the core of QSs for $B_0 = 1\times10^{19}$ G.
Furthermore, for the larger $B_0$ (i.e., $B_0 = 1\times10^{19}$ G), one can see
from Fig.~\ref{Brho} that different B-profiles have obvious effects on $P_\perp$
and $P_\parallel$ as well as their splitting.

\begin{figure}[tbp]
\includegraphics[scale=0.78]{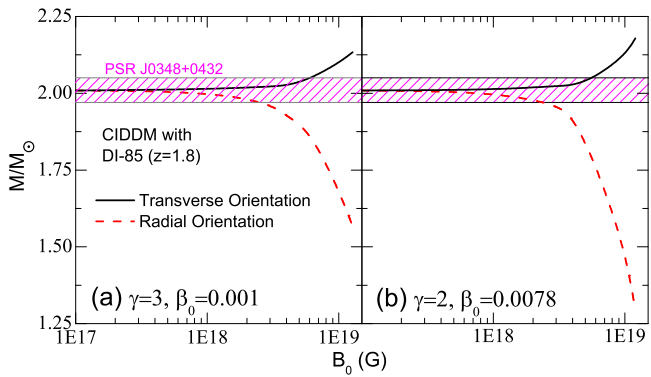}
\caption{(Color online) Maximum mass of static QSs using the
transverse and radial orientations of the magnetic fields as a function
of $B_0$ with the fast B-profile (a) and the slow B-profile (b) within
the CIDDM model with DI-85(z=1.8). The shaded band represents the pulsar
mass of $2.01 \pm 0.04M_{\odot }$ from PSR J0348+0432~\cite{Ant13}.}
\label{MB0}
\end{figure}

The strong pressure anisotropy under strong magnetic fields implies the
orientation of the magnetic fields in QSs should play an important
role on the structure of QSs. Shown in Fig.~\ref{MB0} is the
maximum mass of static QSs using the transverse and radial
orientations of the magnetic fields as a function of $B_0$ with the fast
B-profile and the slow B-profile within the CIDDM model with DI-85 (z=1.8).
It is interesting to see that, while the maximum mass of static QSs
increases with $B_0$ for the transverse orientation, it significantly
decreases with $B_0$ for the radial orientation, especially when $B_0$ is
larger than about $3\times10^{18}$ G. This orientation effect becomes more pronounced
for the slow B-profile.

For the fast B-profile, one can see from Fig.~\ref{MB0} (a) that the
maximum mass of static QSs with the transverse (radial) orientation can
reach about $2.13M_{\odot }$ ($1.57M_{\odot }$) at $B_0 \approx 1.28\times10^{19}$ G
which corresponds to the upper limit of $B_0$ and further increasing $B_0$
would lead to negative $P_\parallel$ in the core of the QSs. In order to
see the effect of the magnetic field orientation on the maximum mass of
QSs, we define the normalized mass asymmetry $\delta_m$ for the maximum
QS mass as
\begin{eqnarray}
\delta_m=\frac{M_{\perp}-M_{||}}{(M_{\perp}+M_{||})/2},
\end{eqnarray}
where $M_{\perp}$ ($M_{||}$) represents the maximum mass of QSs with
transverse (radial) orientation. For the fast B-profile, the largest
mass asymmetry is found to be $\delta_m = 30\%$ at $B_0 = 1.28\times10^{19}$
G from Fig.~\ref{MB0} (a).

In addition, for the slow B-profile, it is seen from Fig.~\ref{MB0} (b)
that the maximum mass of static QSs with the transverse (radial) orientation
can reach about $2.18M_{\odot }$ ($1.29M_{\odot }$) at $B_0 = 1.20\times10^{19}$ G
which corresponds to the upper limit of $B_0$ above which
the negative $P_\parallel$ can appear in the core of QSs, and the corresponding
largest mass asymmetry is $\delta_m = 51\%$ at $B_0 = 1.20\times10^{19}$ G. Therefore,
our results indicate that the maximum mass of magnetized QSs may depend on
both the strength distribution and the orientation of the magnetic fields inside
the stars.

It should be mentioned that the above results and discussions
are based on a special interaction parameter set, i.e., DI-85 (z=1.8),
in the CIDDM model. In order to check the robustness of our conclusions
and see the effects of the isospin dependence of equivalent quark mass
on the properties of SQM under strong magnetic fields, we have further
investigated the case with $D_I = 0$, i.e., the parameter set DI-0 (z=1.8),
for which we have $D_I = 0$ (thus the parameters $\alpha$ and $\beta$
are not involved), $D=26.483$ MeV$\cdot$fm$^{-3z}$, $z=1.8$,
$m_{u0}=m_{d0}=5.5$ MeV and $m_{s0}=80$ MeV, and the maximum
QS mass predicted by DI-0 (z=1.8) without including magnetic fields is
$1.88M_{\odot }$. We find that the parameter set DI-0 (z=1.8) generally
predicts a softer EOS of SQM and smaller values of the maximum QS mass
but almost the same magnetic field effects on the pressure anisotropy
(pressure splitting factor $\delta_p $) and the QS mass asymmetry
$\delta_m$, compared with the parameter set DI-85 (z=1.8). Due to the
softening of the EOS of SQM with DI-0 (z=1.8), the values of the critical
magnetic field strength $B_c$ as shown in Fig.~\ref{PTLConstB} decrease
correspondingly and even disappear (e.g., at $n_B = 3n_0$).
These features imply that our main conclusion about the magnetic field
effects on the properties of QSs remains unchanged even if various
interactions are used.

Furthermore, the high density quark matter might be in a color
superconducting phase~\cite{Alf13}. The possible quark color
superconducting phases mainly include the two-flavor color
superconductor (2SC)~\cite{Alf98,Rus04,Rus05}, the color-flavor-locked
(CFL) phase~\cite{Alf99,Raj01,Lug02,Agr09,Ave11,Pau13}, and the
crystalline color superconductor (CCS)~\cite{Alf01}. In the present
work, for simplicity, we have not considered color superconducting
phases. In recent years, significant progress has been made to
understand the magnetic field effects on the properties of quark color
superconducting phases~\cite{Bla99,Fer05,Fer07,Ferrer10,Pau11,Fer14}.
In particular, by modeling the quark confinement within an effective
bag model, the EOS of the magnetic-color-flavor-locked (MCFL) phase
and the corresponding QS structure under a constant magnetic field
have been investigated in Ref.~\cite{Pau11}. It is nice to see that
the magnetic field effects on the pressure anisotropy and the QS mass
obtained in Ref.~\cite{Pau11} are quite similar with our present
predictions based on the density dependent magnetic field strength
in the CIDDM model. It will be interesting to see how exactly the
quark color superconducting phases affect the properties of SQM and
QSs under strong magnetic fields within the CIDDM model.

\section{Conclusions}
\label{summary}

We have studied the properties of strange quark matter and quark stars under
strong magnetic fields by using the confined isospin- and density-dependent mass
model. The equation of state of strange quark matter under constant magnetic fields
has been calculated self-consistently and the pressure of the system is shown to be
anisotropic along and perpendicular to the magnetic field direction with the former
being generally larger than the latter. The pressure of the system may become
significantly anisotropic when the magnetic field strength is very strong and thus
the properties of magnetized quark stars may significantly depend on the magnetic
field orientation inside the stars.

Using a density-dependent magnetic field profile to simulate the magnetic field
strength distribution in a star, we have studied the properties of static
spherical quark stars by considering two hypothetical cases for the orientation of
the magnetic fields inside the stars, i.e., the radial orientation in which the
local magnetic fields are along the radial direction and the transverse orientation
in which the local magnetic fields are perpendicular to the radial direction but
randomly oriented in the plane perpendicular to the radial direction. Based on
these two extreme cases of the magnetic field orientation, we have demonstrated
that the maximum mass of static magnetized quark stars may significantly depend
on the magnetic field orientation inside the stars, and the magnetic fields
with radial (transverse) orientation can significantly decrease (increase) the
maximum mass of the quark stars. The maximum mass of static magnetized quark
stars has also been found to depend on the details of the density-dependent
magnetic field profile.

Therefore, our present results have shown that besides the strength distribution,
the orientation of the magnetic fields inside the quark stars is also important
for the properties of quark stars under strong magnetic fields. For a fixed
density-dependent magnetic field profile, including the magnetic fields can either
enhance or reduce the maximum mass of static magnetized quark stars, depending
on the magnetic field orientation inside the stars.\\

\section*{ACKNOWLEDGMENTS}

This work was supported in part by the NNSF of China under Grant Nos. 11275125
and 11135011, the Shanghai Rising-Star Program under grant No. 11QH1401100, the
``Shu Guang" project supported by Shanghai Municipal Education Commission
and Shanghai Education Development Foundation, the Program for Professor
of Special Appointment (Eastern Scholar) at Shanghai Institutions of Higher
Learning, the National Basic Research Program of China (973 Program) under Contract
No. 2015CB856900, and the Science and Technology Commission of Shanghai
Municipality (11DZ2260700).

\end{document}